\documentclass[showpacs,preprint,pre]{revtex4}
\usepackage{psfrag,epsfig,amsfonts,amssymb,amsmath,graphicx,slashbox}
\usepackage{dcolumn}

\newcommand{\hr}{{\cal H}}

\newcommand{\RR}{{\mathbb R}}

\newcommand{\nn}{{\cal N}}
\newcommand{\ord}{{\cal O}}

\newcommand{\tr}{\mbox{tr}}
\newcommand{\da}{\Delta_{\! A}}
\newcommand{\pa}{\Delta_{\! P}}

\begin{document}

\title{Typicality of pure states randomly sampled according to 
the Gaussian adjusted projected measure}

\author{Peter Reimann}
\affiliation{Universit\"at Bielefeld, Fakult\"at f\"ur Physik, 33615 Bielefeld, Germany}

\begin{abstract}
Consider a mixed quantum mechanical 
state, describing a statistical ensemble in terms 
of an arbitrary density operator $\rho$ of 
low purity, $\tr\rho^2\ll 1$, 
and yielding the ensemble averaged 
expectation value $\tr(\rho A)$
for any observable $A$.
Assuming that the given statistical ensemble 
$\rho$ is generated by randomly sampling
pure states $|\psi\rangle$ according
to the corresponding so-called Gaussian 
adjusted projected measure
$[$Goldstein et al., J. Stat. Phys. 125, 1197 (2006)$]$, 
the expectation value $\langle\psi|A|\psi\rangle$
is shown to be extremely close 
to the ensemble average $\tr(\rho A)$
for the overwhelming majority of pure
states $|\psi\rangle$ and any experimentally 
realistic observable $A$.
In particular, such a `typicality' property holds 
whenever the Hilbert space $\hr$ of the system
contains a high dimensional subspace $\hr_+\subset\hr$ with 
the property that all $|\psi\rangle\in\hr_+$ are 
realized with equal probability and all 
other $|\psi\rangle \in\hr$ are excluded.
\end{abstract}


\maketitle

\section{Introduction}
Spheres in high dimensional Euclidean spaces exhibit
astonishing geometrical properties, as discussed in
detail e.g. in basic Statistical Physics lectures:
two randomly drawn vectors, each connecting the center
of the sphere with any point at its surface, are practically
orthogonal with extremely high probability;
almost the entire volume of the sphere is contained 
within an extremely thin surface layer of the sphere;
the latter in turn exhibits an extreme concentration of
its volume around a very narrow `equatorial belt',
and so on.
In quantum mechanics, pure states live on unit
spheres in Hilbert spaces of usually very high 
dimension,
and hence one naturally may wonder about their 
corresponding peculiarities.
One of them is the subject of our present paper.
Namely, we will show the following main result:
Consider a mixed state, describing a statistical 
ensemble in terms of a density operator $\rho$
with low purity, $\tr\rho^2\ll 1$, meaning that
the mixed state is very `far' from 
resembling any pure state.
Yet, the statistical ensemble $\rho$ can be
thought of as arising by randomly sampling
pure states $|\psi\rangle$ according
to some probability distribution.
In fact, it is well known (see Sect. II for 
more details), that there are many different 
probability distributions of pure states 
$|\psi\rangle$ which give rise to the same 
mixed state $\rho$.
Here we show that for any given $\rho$
of low purity there exists at 
least one such probability distribution
with the following quite astonishing 
property:
Given an observable $A$, the
expectation value $\langle\psi|A|\psi\rangle$
for the overwhelming majority of pure
states $|\psi\rangle$ is extremely close 
to the ensemble averaged expectation value 
$\tr(\rho A)$ compared to the
full range of {\em a priori}
possible expectation values
$\max_{|\psi\rangle} \langle\psi|A|\psi\rangle-
\min_{|\psi\rangle} \langle\psi|A|\psi\rangle$
(the latter difference is tacitly assumed to 
be finite, as is the case for any 
experimentally realistic observable $A$, 
see Sect. III).

For this kind of property, the term `typicality' 
has been coined in \cite{gol06a}.
While such `typicality' results are applicable
in principle to general quantum mechanical systems, 
they are obviously 
of particular interest with respect to the
foundation of statistical physics of macroscopic
systems at equilibrium, as discussed in
detail e.g. in Refs. 
\cite{pop06,llo06,gol06a,gol06b,rei07,sug07}.
Further related works include
\cite{llo88,gem04,simulations,early}.
With our present study we extend previous 
results from \cite{pop06,gol06a,gol06b,rei07,sug07}
to yet another important class of 
probability distributions of the 
pure states $|\psi\rangle$,
namely the so-called Gaussian adjusted 
projected measure (GAP), 
recently introduced in Ref. \cite{gol06b}.

\section{Outline of the problem}
We consider a quantum mechanical system with 
(separable) Hilbert space $\hr$ of dimension $N\leq\infty$.
The system is assumed to be in a mixed state 
(statistical ensemble) described by
a density matrix $\rho$.
Let $\{|n\rangle\}_{n=1}^N$ be an orthonormal basis of 
eigenvectors of $\rho$ and $p_n$ the corresponding 
eigenvalues,
\begin{equation}
\rho=\sum_{n=1}^N p_n|n\rangle\langle n| 
\label{1a}
\end{equation}
with the usual properties 
\begin{eqnarray}
p_n & \geq & 0
\label{1b}
\\
\sum_{n=1}^N p_n & = & 1 \ 
\label{1c}
\end{eqnarray}
In the context of equilibrium statistical mechanics,
$|n\rangle$ will usually be the eigenstates of the
system Hamiltonian, but we will not make use
of such a property anywhere in this paper.
A particularly simple and important example is the
{\em microcanonical density operator} 
with
\begin{equation}
\rho_{mic}=\frac{1}{N_+}\sum_{n\in S} |n\rangle\langle n| 
\label{1d}
\end{equation}
or, equivalently, with
\begin{equation}
p_n = 1/N_+\ \mbox{if } n\in S,
\ p_n = 0\ \mbox{if } n\not\in S \ .
\label{1e}
\end{equation}
where $S$ is a subset of $\{1,...,N\}$, consisting
of $N_+$ elements ($1\leq N_+\leq N)$.

Given a density matrix of the general form (\ref{1a})-(\ref{1c}),
an arbitrary, normalized pure state (e.g. a wave function) 
can be written in the form
\begin{equation}
|\psi\rangle=\sum_{n=1}^N z_n |n\rangle
\label{4a}
\end{equation}
where $z_n:=\langle n|\psi\rangle$ are complex
coefficients, satisfying the normalization condition
\begin{equation}
||{\bf z}||=1 
\label{5a}
\end{equation}
with the standard definitions
\begin{eqnarray}
{\bf z} & := & (z_1,z_2, ... , z_N)
\label{6a}
\\
||{\bf z}|| & := & \left(\sum_{n=1}^N |z_n|^2\right)^{1/2} \ .
\label{11a}
\end{eqnarray}

Next, we assume that the statistical ensemble $\rho$
is generated by randomly sampling pure states 
(\ref{4a}) according to some probability density 
$p({\bf z})$.
The corresponding ensemble average of an arbitrary 
function $f({\bf z})$ is denoted by
\begin{equation}
\overline{f({\bf z})}:=\int d{\bf z}\, f({\bf z})\, p({\bf z}) \ ,
\label{14a}
\end{equation}
where $d{\bf z}$ represents the natural, uniform measure 
for the $N$-dimensional complex argument ${\bf z}$,
\begin{equation}
d{\bf z} := \prod_{n=1}^N d(\mbox{Re} z_n) \  d(\mbox{Im} z_n) 
\label{10b}
\end{equation}
For infinite dimensional systems, well defined limits
$N\to\infty$ are tacitly taken for granted 
in (\ref{14a}) and in similar
expressions later on.

Put differently, by averaging pure states 
(\ref{4a}), represented 
as projectors $|\psi\rangle\langle \psi|$, according
to the probability density $p({\bf z})$, the given
statistical ensemble $\rho$ has to be reproduced, i.e.
\begin{equation}
\overline{|\psi\rangle\langle\psi|} = \sum_{m,n=1}^N \overline{z_m^\ast z_n}
\ |n\rangle\langle m| = \sum_{n=1}^N p_n \ |n\rangle\langle n|
=\rho
\label{18a}
\end{equation}
where the star indicates complex conjugation
and where we exploited (\ref{1a}) in the last identity. 
Hence, the second moments of the
distribution $p({\bf z})$ are fixed by the given 
statistical ensemble $\rho$ via 
\begin{equation}
\overline{z_m^\ast z_n} = \delta_{mn} \, p_n  \ .
\label{18b}
\end{equation}
In turn, every $p({\bf z})$ with  second moments (\ref{18b})
reproduces the given $\rho$ in (\ref{18a}).
We thus recover the well known fact that 
{\em a given density matrix $\rho$ does not 
uniquely fix the distribution of pure states 
$p({\bf z})$}. 
For more details, explicit examples, and
further references see e.g. Ref. \cite{gol06b}.

Next we consider an arbitrary observable 
$A=A^\dagger:\hr\to\hr$ with eigenvectors 
$|\nu\rangle$ and eigenvalues $a_\nu$, i.e.
\begin{equation}
A=\sum_{\nu=1}^N a_\nu |\nu\rangle\langle\nu| \ .
\label{2b}
\end{equation}
Here and in the following we use the convention that 
Greek labels $\nu$ and $\mu$ implicitly refer to the 
eigenvectors of $A$, which is convenient but somewhat 
ambiguous in so far as, e.g., $|\nu =3\rangle$ is 
not the same vector as $|n=3\rangle$.
According to (\ref{1a}), the ensemble averaged 
expectation value of $A$ is given by
\begin{eqnarray}
\langle A\rangle & := &\tr(\rho A)=\sum_{n=1}^N p_n A_{nn} 
\label{2a}
\\
A_{mn} & := & \langle m|A|n\rangle \ .
\label{3a}
\end{eqnarray}
Further, any given pure state $|\psi\rangle$ gives rise to 
an expectation value $\langle\psi|A|\psi\rangle\in\RR$.
The random distribution of those expectation 
values, induced by the distribution $p({\bf z})$ of 
pure states, is the quantity of central
interest in our present work.
In particular, we will be interested in 
identifying conditions under which the
expectation values $\langle\psi|A|\psi\rangle$
will be very close to the ensemble average (\ref{2a})
with very high probability, 
i.e. for a large majority of pure states $|\psi\rangle$.

\section{Relevant Hilbert space and Observables}
Without loss of generality, we assume that the indices $n$\
in (\ref{1a}) are ordered so that
\begin{equation}
p_n>0\ \mbox{for}\ n\leq N_+ \ \mbox{and}\  p_n=0\ \mbox{for}\ n>N_+
\label{7a}
\end{equation}
for some integer $N_+$ with $1\leq N_+\leq N$.
The $N_+$-dimensional sub Hilbert space spanned by the 
basis vectors $\{|n\rangle\}_{n=1}^{N_+}$ 
is denoted by $\hr_+$ and the projector
onto this subspace by 
\begin{equation}
P_+:=\sum_{n=1}^{N_+} |n\rangle\langle n| \ .
\label{8a}
\end{equation}
In particular, $P_+$ is at the same time the identity operator
on $\hr_+$ and the microcanonical density operator
(\ref{1d}) takes the form
\begin{equation}
\rho_{mic}=P_+/N_+ \ .
\label{8b}
\end{equation}

From (\ref{18b}) and (\ref{7a}) we conclude \cite{gol06b}
that (with probability one) $z_n=0$ for $n>N_+$ and hence
$|\psi\rangle\in\hr_+$ according to (\ref{4a}).
As pointed out in Sect. II, our main goal is to
determine the probability distribution of the 
expectation values $\langle\psi|A|\psi\rangle$. 
It follows that with respect to this goal only
the restriction/projection 
\begin{equation}
A_+:=P_+AP_+
\label{8c}
\end{equation}
of the observable $A$ to the subspace $\hr_+$ is relevant.
Equivalently, {\em whenever $m$ or $n$ exceeds $N_+$
then the matrix element $A_{mn}$ is of no relevance
for our purpose} and thus can be set to zero 
without loss of generality.

The full range of possible expectation values
$\langle\psi|A|\psi\rangle$ is quantified by
\begin{eqnarray}
\da & := & 
\max_{|\psi\rangle\in\hr_+} \langle\psi|A_+|\psi\rangle
-\min_{|\psi\rangle\in\hr_+} \langle\psi|A_+|\psi\rangle
\nonumber
\\
& = & 
\max_{|\psi\rangle\in\hr_+} \langle\psi|A|\psi\rangle
-\min_{|\psi\rangle\in\hr_+} \langle\psi|A|\psi\rangle
\nonumber
\\
& \leq & 
\max_{|\psi\rangle\in\hr} \langle\psi|A|\psi\rangle
-\min_{|\psi\rangle\in\hr} \langle\psi|A|\psi\rangle \ .
\label{8d}
\end{eqnarray}
The second relation follows from (\ref{8c}) and the fact
that $P_+$ is the projector onto the subspace $\hr_+$
and the last relation from the fact that $\hr_+\subset\hr$.
Alternatively, $\da$ can thus be identified with the
difference between the largest and the smallest eigenvalues
of $A_+$ and is bounded from above by the difference
between the largest and the smallest eigenvalues of $A$,
cf. Eq. (\ref{2b}).

Clearly, any given real experimental apparatus has 
a finite range and hence the corresponding
range of possible expectation values 
$\da$ from Eq. (\ref{8d}) is finite.
This range is specific to the given measurement device,
but is (practically) independent of the properties
(e.g. the size) of the observed system.
Here and in the following we restrict ourselves
to observables $A$ in the above sense.
For instance, the energy of a harmonic oscillator
is not an observable in this sense:
in principle, the energy of the oscillator
may become arbitrarily large (albeit with extremely
small probability), but no real device would be
able to display its value beyond a certain upper
limit. Rather, all energies beyond this limit will
yield one and the same measurement result
(e.g. a blow up of the device), and hence 
only the corresponding `truncated' energy 
operator would be an admissible `observable'.

We emphasize again that the above restriction
regarding the admissible observables $A$ is
of a purely mathematical/formal nature,
it does not exclude any relevant observable
corresponding to a realistic physical
measurement.
Indeed, it is well known that any realistic
observable can be built up by means of
suitable projector operators, and each
such projector $P$ only has eigenvalues
zero and unity and hence $\pa =1$.
Essentially, the same `restriction' in fact
also applies to the seminal prior works \cite{gol06a,pop06,sug07}.
The measure of distance employed in \cite{pop06} 
is the trace-norm, involving a maximization over
all operators with operator norm bounded by unity.
Hence, the estimates from \cite{pop06} become worse 
and worse, as the maximally admissible 
norm of the considered operators increases. 
Similar conclusions apply for the estimates 
from \cite{gol06a}. 
The most explicit discussion of this issue is contained 
in \cite{sug07}.

To summarize, only the sub Hilbert space
$\hr_+$ and the projected observables (\ref{8c})
are of relevance for our purpose.
Accordingly, we can and will for simplicity 
assume in some of the 
following sections temporarily
that $\hr=\hr_+$ and thus $A=A_+$.
In other words, all subscripts `$+$' 
will be omitted and it will be taken 
for granted that $p_n>0$ for all $n$.
Furthermore, we can and will focus on
observables of finite range $\da$ 
according to (\ref{8d}).

\section{The Gaussian adjusted projected measure (GAP)}
To avoid unnecessary technical complications, 
we temporarily restrict ourselves to finite 
dimensions $N$. 
In the final results of our calculations,
the limit $N\to\infty$ can be readily performed.
Furthermore, we assume $p_n>0$ for all $n$ without
loss of generality, see end of Sect. III.

Taking for granted  the above assumptions that $N<\infty$ and $p_n>0$, we define
\begin{eqnarray}
p({\bf z}) & := & \int d{\bf y}\, 
\nn \exp\left(-\sum_{n=1}^N \frac{|y_n|^2}{p_n}\right)\,
||{\bf y}||^2\,\delta({\bf z}-{\bf y}/||{\bf y|}|) \ .
\label{10a}
\end{eqnarray}
Here, $d{\bf y}$ is defined like in (\ref{10b}), i.e. we are dealing with
an integral over $2N$ real integration variables, and 
$\nn:=\prod_{n=1}^N (\pi p_n)^{-1}$ is a normalization constant 
(see below).
The delta-function is by definition understood in such a way that
the ensemble average of an arbitrary function $f({\bf z})$ 
from (\ref{14a}) takes the form
\begin{eqnarray}
\overline{f({\bf z})} & = &
\int d{\bf y}\, \nn \exp\left(-\sum_{n=1}^N \frac{|y_n|^2}{p_n}\right)\,
||{\bf y}||^2\,  f({\bf y}/||{\bf y|}|) 
\label{15a1}
\end{eqnarray}

Eq. (\ref{10a}) defines the Gaussian adjusted projected (GAP) measure 
\cite{gol06b} associated with the density matrix (\ref{1a}), 
written in the form of a probability density $p({\bf z})$ 
with respect to the natural measure (\ref{10b}).
The word `Gaussian' in the notion GAP refers \cite{gol06b} to 
the exponential factor in (\ref{10a});
the word `adjusted' refers to the factor $||{\bf y}||^2$,
which is needed to fulfill condition (\ref{18a}) (see below);
the word `projected' refers to the delta-function in 
(\ref{10a}), guaranteeing the normalization condition 
(\ref{5a}) (see below).

In the special case of a microcanonical density
operator we have $p_n=1/N$ for all $n$ according
to (\ref{1e}) and our assumption at the beginning of this 
section, yielding with (\ref{15a1}) the result
\begin{eqnarray}
\overline{f({\bf z})} & = &
\int d{\bf y}\, \nn\, e^{-||{\bf y}||^2 N} 
||{\bf y}||^2\,  f({\bf y}/||{\bf y}||) \ .
\label{15a2}
\end{eqnarray}
It follows that $\overline{f(U{\bf z})}=\overline{f({\bf z})}$
for arbitrary unitary $N\times N$ matrices $U$.
Since also $f({\bf z})$ is arbitrary, we recover the
fact \cite{gol06b} that two arguments ${\bf z}$ with equal 
length $||{\bf z}||$ are realized with equal probability.

Returning to the general case, it is often convenient to 
change from a Cartesian representation of the 
complex numbers $y_n$
in terms of real and imaginary parts (cf. (\ref{10b}))
to a polar representation in terms of 
$r_n\geq 0$ and $\varphi_n\in [0,2\pi)$ 
via the usual relation $y_n=r_n e^{i\varphi_n}$.
Then, the ensemble average of an arbitrary function $f({\bf z})$ 
from (\ref{15a1}) can be rewritten as
\begin{eqnarray}
\overline{f({\bf z})} & = &
\left[\prod_{n=1}^N\int_0^\infty dr_n\int_0^{2\pi} d\varphi_n
\frac{r_n}{\pi p_n}\,e^{-r_n^2/p_n}\right] 
\, ||{\bf r}||^2\, f({\bf c}) 
\label{15a}
\\
c_n & := & r_n e^{i\varphi_n}/||r|| \ .
\label{17a}
\end{eqnarray}

We first consider the special choice $f({\bf z}):=1$,
implying with (\ref{14a}) and (\ref{15a}) that 
\begin{equation}
\int  d{\bf z}\,  p({\bf z}) =
\left[\prod_{l=1}^N\int_0^\infty dr_l\int_0^{2\pi} d\varphi_l
\frac{r_l}{\pi p_l}\,e^{-r_l^2/p_l}\right] 
\, \sum_{n=1}^N r_n^2 \ .
\label{17b}
\end{equation}
The $N$ integrals over $\varphi_l$ are trivial,
each yielding a factor $2\pi$.
Hence, we can infer that
\begin{equation}
\int  d{\bf z}\, p({\bf z}) =
\sum_{n=1}^N 
\prod_{l=1}^N\int_0^\infty dr_l \frac{2r_l^{1+2\delta_{ln}}}{p_l}\,e^{-r_l^2/p_l} \ .
\label{17c}
\end{equation}
The integrals over $r_l$ for $l\not=n$ are of the form
\begin{equation}
\int_0^\infty dr_l \frac{2r_l}{p_l}\,e^{-r_l^2/p_l}
= \int_0^\infty dr_l \, \left(-\frac{d}{dr_l}\right)e^{-r_l^2/p_l} = 1 \ .
\label{17c1}
\end{equation}
Likewise, the integral over $r_l$ for $l=n$ is of the form
\begin{equation}
\int_0^\infty dr_l \frac{2r_l^3}{p_l}\,e^{-r_l^2/p_l} 
= \int_0^\infty dr_l \, r_l^2\left(-\frac{d}{dr_l}\right)e^{-r_l^2/p_l} 
= \int_0^\infty dr_l \, 2 r_l e^{-r_l^2/p_l} = p_l
\label{17c2}
\end{equation}
where the second identity follows by a partial integration and
the last identity by means of (\ref{17c1}).
All in all, the right hand side of (\ref{17b}) thus amounts
to $\sum_{n=1}^N p_n$ and with (\ref{1c}) we see that
$p({\bf z})$ is normalized to unity.
Observing that the right hand side in (\ref{10a})
is non-negative for any ${\bf z}$,
we can conclude that $p({\bf z})$ 
is indeed a well-defined probability density.

Next, we consider the special choice 
$f({\bf z}):=\delta\left(X-\sum_{n=1}^N |z_n|^2\right)$ for an 
arbitrary real number $X$.
According to (\ref{17a}), the argument $f({\bf c})$ in (\ref{15a})
takes the form $\delta(X-1)$ and thus can be brought
in front of all the integrals. 
The remaining integral is identical to the one 
evaluated in the preceding paragraph, 
i.e. its value is unity, and hence
$\overline{f({\bf z})}=\delta(X-1)$.
It follows that $p({\bf z})$ indeed takes non-zero values only for 
arguments ${\bf z}$ respecting the normalization condition (\ref{5a}).

Finally, we consider the special choice $f({\bf z}):=z_m^\ast z_n$.
Exploiting (\ref{15a}) we obtain
\begin{eqnarray}
& & \overline{z_m^\ast z_n}=
\left[\prod_{l=1}^N\int_0^\infty dr_l\int_0^{2\pi} d\varphi_l
\, \frac{r_l}{\pi p_l}\,e^{-r_l^2/p_l}\right] 
\, r_mr_n e^{i(-\varphi_m+\varphi_n)} \ .
\label{17d}
\end{eqnarray}
If $m\not=n$, the integral 
over $\varphi_n$ can be carried out first,
being proportional to $\int_0^{2\pi} d\varphi_n
\, e^{i\varphi_n}=0$. 
Hence $\overline{z_m^\ast z_n}=0$ if $m\not=n$.
In the case $m=n$ we have 
$e^{i(-\varphi_m+\varphi_n)}=1$
and we are left with $N$ independent
integrals of the form 
$\int_0^{2\pi} d\varphi=2\pi$.
The remaining integrals over $r_l$ are of the
same type as those already encountered in
(\ref{17c1}) and (\ref{17c2}) yielding the final result
$\overline{z_m^\ast z_n}=p_n$.
All together we thus find that the GAP
measure (\ref{10a}) indeed fulfills
the condition (\ref{18b}) and hence 
reproduces  the correct 
statistical ensemble (\ref{18a}) 
encoded by the preset density operator 
$\rho$ from (\ref{1a}).
Without the `adjusting factor' $||{\bf r}||^2$ in (\ref{10a})
this property could not be maintained \cite{gol06b}.

\section{Evaluation of the variance}
As pointed out in Sect. II, our main goal is to
determine the probability distribution of the 
expectation values $\langle\psi|A|\psi\rangle$
induced by the distribution 
$p({\bf z})$ of pure states according to the GAP measure (\ref{10a}).
For the first moment, $\overline{\langle\psi|A|\psi\rangle}$, 
the expected result $\langle A\rangle$ is readily recovered
by means of (\ref{18a}) and (\ref{2a}):
\begin{equation}
\overline{\langle\psi|A|\psi\rangle}
=\overline{\tr(|\psi\rangle\langle\psi|A)}
=\tr(\overline{|\psi\rangle\langle\psi|}A)
=\tr(\rho A)=\sum_{n=1}^N p_n\, A_{nn}=\langle A\rangle \ .
\label{20a}
\end{equation}
In the present Section, our focus is on the variance
\begin{equation}
\sigma_{\!\! A}^2 :=
\overline{[\langle\psi|A|\psi\rangle - \langle A \rangle]^2} 
=\overline{\langle\psi|A|\psi\rangle^2} -\overline{\langle\psi|A|\psi\rangle}^2 \ .
\label{20b}
\end{equation}
We emphasize, that this variance characterizes the dispersion of
the expectation value of $A$ for different pure states $|\psi\rangle$,
and not the ``quantum fluctuations'' associated with individual 
measurements of $A$ of a fixed pure state $|\psi\rangle$.

Observing that the two observables $A$ and $A-\langle A \rangle$
have the same variance and the same range $\da$ according
to (\ref{8d}), we can and will restrict ourselves in
this section without loss of generality to observables $A$
with the  property
\begin{equation}
\sum_{n=1}^N p_n\, A_{nn}=\langle A\rangle = 0 \ .
\label{20c}
\end{equation}
Furthermore, we maintain the assumptions $N<\infty$ and
$p_n>0$ for all $n$, as introduced at the beginning of
the previous Section. In particular, we thus have $A=A_+$ and 
both the eigenvalues $a_\nu$ (cf. (\ref{2b})) and the
diagonal matrix elements $A_{nn}$ (cf. (\ref{3a}))
are bounded from above by 
$a_{max}:=\max_{|\psi\rangle\in\hr} \langle\psi|A|\psi\rangle=\max_\nu a_\nu$
and from below by
$a_{min}:=\min_{|\psi\rangle\in\hr} \langle\psi|A|\psi\rangle=\min_\nu a_\nu$.
In view of (\ref{20c}) it follows that $a_{max}\geq 0$ and
$a_{min}\leq 0$ and hence with (\ref{8d}) that
\begin{equation}
|A_{nn}|,\, |a_\nu|\leq\da\ \mbox{for all}\ n,\,\nu \ .
\label{20d}
\end{equation}

With of (\ref{4a}) and (\ref{20c}), the variance (\ref{20b})
takes the form
\begin{eqnarray}
\sigma_{\!\! A}^2
& = & \overline{\left[\sum_{m,n} z_m^\ast z_n A_{mn}\right]^2}
=\sum_{j,k=1}^N\sum_{m,n=1}^N A_{jk}A_{mn} 
\overline{ z_j^\ast z_k  z_m^\ast z_n}
\label{21a}
\end{eqnarray}
The average in the last term can be rewritten by means of
(\ref{15a}) as
\begin{eqnarray}
& & \overline{ z_j^\ast z_k  z_m^\ast z_n}=
\left[\prod_{l=1}^N\int_0^\infty dr_l\int_0^{2\pi} d\varphi_l
\, \frac{r_l}{\pi p_l}\,e^{-r_l^2/p_l}\right] 
\, \frac{r_jr_kr_mr_n e^{i(-\varphi_j+\varphi_k-\varphi_m+\varphi_n)}}{||{\bf r}||^2} \ .
\label{22a}
\end{eqnarray}
The evaluation of these integrals is analogous but
somewhat more involved than those from the preceding
Section:
The integrals over the angles $\varphi_l$ can be readily performed,
yielding a factor of $(2\pi)^N$ in the two cases
(i) $j=k$ and $m=n$, (ii) $j=n$ and $k=m$,
and zero in any other case.
Taking care not to count the case $j=k=m=n$ twice
and after a convenient renaming of the 
summation indices we thus obtain
\begin{eqnarray}
\sigma_{\!\! A}^2
& = &
\sum_{m\not = n} [A_{mm}A_{nn}+ A_{mn}A_{nm}] I_{mn} 
+\sum_{n} [A_{nn}]^2 I_{nn} 
\label{23a}
\\
I_{mn}& := &
\left[\prod_{l=1}^N\int_0^\infty dr_l\, \frac{2 r_l}{p_l}\,e^{-r_l^2/p_l}\right] 
\, \frac{r_m^2r_n^2}{||{\bf r}||^2} \ .
\label{24a}
\end{eqnarray}

In order to evaluate the integral $I_{mn}$, we consider the auxiliary
function
\begin{equation}
h(x ,{\bf y})  :=  \left[\prod_{l=1}^N\int_0^\infty 
dr_l\, \frac{2 r_l}{p_l}\right] \ 
\exp\left\{-\sum_{l=1}^N(x +y_l)r_l^2\right\}
\label{25a}
\end{equation}
where $x \geq 0$, ${\bf y}:=(y_1,....,y_N)$, 
and $y_n>0$ for all $n$.
Observing that the right hand side in (\ref{25a}) factorizes into 
$N$ independent integrals of the form $\int_0^\infty dr_l\, r_l\, e^{-br_l^2}=1/2b$ 
with $b:=x +y_l>0$ (see also (\ref{17c1})), we obtain
\begin{equation}
h(x ,{\bf y})=\prod_{l=1}^N\frac{1}{p_l}\,\frac{1}{x +y_l} \ .
\label{26a}
\end{equation}
Next, we note that the integral over the $x $-dependent terms in
(\ref{25a}) is of the form $\int_0^\infty dx \, e^{-x  ||{\bf r}||^2}=1/||{\bf r}||^2$
(see (\ref{11a})), implying that
\begin{equation}
H({\bf y}):=\int_0^\infty dx \, h(x ,{\bf y})=
\left[\prod_{l=1}^N\int_0^\infty 
dr_l\, \frac{2 r_l}{p_l}\right] \ 
\exp\left\{-\sum_{l=1}^N y_l r_l^2\right\}/||{\bf r}||^2 \ .
\label{27a}
\end{equation}
By comparison with (\ref{24a}) we can conclude that
\begin{equation}
I_{mn}=\frac{\partial^2 H({\bf y})}{\partial y_m\partial y_n}\bigg|_{y_l=1/p_l}
\label{28a}
\end{equation}
By combining (\ref{26a})-(\ref{28a}) it follows that
\begin{eqnarray}
I_{mn} & = & \int_0^\infty dx \, 
\frac{\partial^2}{\partial y_m\partial y_n}
\prod_{l=1}^N\frac{1}{p_l}\,\frac{1}{x +y_l}\bigg|_{y_l=1/p_l}
=p_mp_n(1+\delta_{mn})K_{mn}
\label{29a}
\\
K_{mn} & := & \int_0^\infty dx \, g_{mn}(x)\, G(x)
\label{30a}
\\
g_{nm}(x) & := & (1+xp_m)^{-1}(1+xp_n)^{-1}
\label{30b}
\\
G(x) & := & \prod_{l=1}^N (1+xp_l)^{-1}
\label{30c}
\end{eqnarray}
Finally, this yields for the variance (\ref{23a}) the result
\begin{eqnarray}
\sigma_{\!\! A}^2
& = &
\sum_{m,n=1}^N [A_{mm}A_{nn}+ A_{mn}A_{nm}] p_mp_n K_{mn} \ .
\label{31a}
\end{eqnarray}

Next we turn to a more detailed discussion of 
the integrals $K_{mn}$ in (\ref{30a}). 
Clearly, the integrand is a positive function of $x$,
bounded from above by unity, and decaying like 
$1/x^{N+2}$ for large $x$ due to our assumption
that $p_n>0$ for all $n$, see below (\ref{20c}).
Hence the integrals $K_{mn}$ are finite
and positive.
Specifically, for the microcanonical density
operator we have $p_n=1/N$ for all $n$ according
to (\ref{1e}) and our assumption below 
(\ref{20c}), yielding with (\ref{30a}) the exact result
\begin{eqnarray}
K_{mn} & = & \int_0^\infty dx \ (1+x/N)^{-N-2}
= \frac{N}{N+1} \ . 
\label{32a}
\end{eqnarray}

To further evaluate $K_{mn}$ in the general case,
we rewrite $g_{mn}(x)$ from (\ref{30b}) by means of
Taylor's theorem \cite{taylor} as
\begin{eqnarray}
g_{mn} (x) & = & g_{mn} (0)+ x g_{mn}' (0)+ \frac{x^2}{2} g_{mn}'' (x\theta_{mn}(x))
\nonumber
\\
& = &  1 - x \, (p_m+p_n) + x^2\, (p_m^2+p_mp_n+p_n^2)\, \chi_{mn}(x)
\label{50a}
\end{eqnarray}
for certain functions $\theta_{mn}(x)$ and
$\chi_{mn}(x)$, satisfying 
$\theta_{mn}(x)$, $\chi_{mn}(x)\in [0,1]$ 
for all $x\geq 0$, but for the rest
depending in a non-trivial manner on $x$, $m$, and $n$.
Note that while an infinite power series expansion 
would not converge for arbitrary $x\geq 0$,
the above finite order Taylor expansion  
is an exact identity \cite{taylor} for all $x\geq 0$.
As a consequence, (\ref{30a}) can be rewritten as
\begin{eqnarray}
K_{mn} & = & K^{(0)}-(p_m+p_n)\, K^{(1)} + 
2\, (p_m^2+p_mp_n+p_n^2) \, \kappa_{mn}\, K^{(2)}
\label{51a}
\\
K^{(k)} & := & \frac{1}{k!} \int_0^\infty dx \, x^k\, G(x), \ k=0,1,2
\label{52a}
\\
\kappa_{mn} & \in & [0,1]
\label{53a}
\end{eqnarray}
From (\ref{30c}) and $p_n>0$ for all $n$ we can infer
that the integrals in ({\ref{52a}) are finite 
(and positive) if and only if 
\begin{equation}
N\geq 4 \ .
\label{53b}
\end{equation}
The latter condition is tacitly taken for granted
henceforth.

Next we return to the variance in (\ref{31a}). 
In view of (\ref{20c}) we see that
sums of the form
\begin{eqnarray}
\sum_{m,n=1}^N A_{mm}A_{nn} p_mp_n Q_{mn} 
=\sum_{m=1}^N A_{mm}p_m\sum_{n=1}^N A_{nn}p_n Q_{mn}
\label{53c}
\end{eqnarray}
are zero if the coefficients $Q_{mn}$ are either independent
of $m$ or independent of $n$. 
Hence the first sum on the right 
hand side of (\ref{31a}) vanishes in the special case
(\ref{32a}) corresponding to the microcanonical density operator.
Likewise, in the general case we can conclude
with (\ref{51a}) that the first sum on the right 
hand side of (\ref{31a}) takes the form
\begin{eqnarray}
\sum_{m,n=1}^N A_{mm}A_{nn} p_mp_n K_{mn}
=
2 K^{(2)} \sum_{m,n=1}^N A_{mm}A_{nn} p_mp_n 
(p_m^2+p_mp_n+p_n^2) \, \kappa_{mn} \ ,
\label{53d}
\end{eqnarray}
yielding with (\ref{53a}) and $K^{(2)}\geq 0$ the
estimate
\begin{eqnarray}
|\sum_{m,n=1}^N A_{mm}A_{nn} p_mp_n K_{mn}|
\leq
2 K^{(2)} \sum_{m,n=1}^N |A_{mm}| \, |A_{nn}| (2 p_mp^3_n+ p^2_mp^2_n) \ .
\label{53e0}
\end{eqnarray}
With (\ref{1c}) and (\ref{20d}) we obtain
\begin{eqnarray}
\sum_{m,n=1}^N A_{mm}A_{nn} p_mp_n K_{mn}
\leq
2 K^{(2)} \da^2 \left(2\sum_{n=1}^N p^3_n+ \left[\sum_{n=1}^N p^2_n\right]^2\right) \ .
\label{53e}
\end{eqnarray}
Turning to the second sum on the right hand side of (\ref{31a}), 
we note that
\begin{eqnarray}
0\leq  \sum_{m,n=1}^N A_{mn}A_{nm} p_mp_n K_{mn}
\leq K^{(0)} \sum_{m,n=1}^N A_{mn}A_{nm} p_mp_n  \ .
\label{53g}
\end{eqnarray}
The first inequality follows from the fact that
$A_{mn}A_{nm}=|A_{mn}|^2\geq 0$, $p_m p_n\geq 0$,
and $K_{mn}\geq 0$ for all $m$, $n$.
The second inequality follows from $g_{mn}(x)\leq 1$ according
to (\ref{30b}), hence $K_{mn}\leq K^{(0)}$ according to
(\ref{30a}) and (\ref{52a}).
By means of (\ref{1a}) and (\ref{3a}) one readily finds that
\begin{equation}
\tr(\rho A)^2 =
\sum_{m=1}^N \langle m|\rho A \sum_{n=1}^N |n\rangle\langle n| \rho A|m\rangle
= \sum_{m,n=1}^N p_m \langle m|A|n\rangle p_n \langle n|A|m\rangle
= \sum_{m,n=1}^N p_mp_n A_{mn}A_{nm} \ .
\label{53i}
\end{equation}
Likewise, by using the eigenvectors $|\nu\rangle$
and eigenvalues $a_\nu$ of $A$ from (\ref{2b}) to evaluate the trace 
one obtains
\begin{equation}
\tr(\rho A)^2 =
\sum_{\mu=1}^N \langle \mu|\rho A \sum_{\nu=1}^N |\nu \rangle\langle \nu| 
\rho A|\mu\rangle
= \sum_{\mu,\nu=1}^N \langle \mu|\rho|\nu \rangle a_\nu 
\langle \nu|\rho |\mu \rangle a_\mu
=\sum_{\mu,\nu=1}^N a_\mu a_\nu \rho_{\mu\nu}\rho_{\nu\mu} \ .
\label{53j}
\end{equation}
Combining (\ref{53g})-(\ref{53j}) and
$\rho_{\mu\nu}\rho_{\nu\mu}=|\rho_{\mu\nu}|^2$ yields
\begin{eqnarray}
\sum_{m,n=1}^N A_{mn}A_{mn} p_mp_n K_{mn}
\leq K^{(0)} \sum_{\mu,\nu=1}^N a_\nu a_\mu |\rho_{\mu\nu}|^2 
\leq K^{(0)} \da^2 \sum_{\mu,\nu=1}^N  |\rho_{\mu\nu}|^2 
= K^{(0)} \da^2 \, \tr \rho^2 \ ,
\label{53k}
\end{eqnarray}
where the second inequality follows from (\ref{20d}) and the last
equality from (\ref{53j}) with $A=1$.

In the special of a microcanonical density operator 
(\ref{1d}) we have seen below (\ref{53c}) that the 
first sum on 
the right hand side of (\ref{31a}) vanishes. 
Exploiting (\ref{32a}), (\ref{53i}), and 
the fact that $p_n=1/N$ for all $n$ we obtain
\begin{equation}
\sigma_{\!\! A}^2 = \frac{N}{N+1}\tr(\rho_{mic} A)^2 =
\frac{\tr A^2}{N(N+1)} = \frac{\sum\limits_{\nu=1}^N a_\nu^2}{N^2}\, 
\left[1+\ord\left(\frac{1}{N}\right)\right]
\ ,
\label{53k1}
\end{equation}
where $a_\nu$ are the eigenvalues of $A$, see (\ref{2b}).

Returning to the general case, the variance (\ref{31a}) can be
estimated from above by means of (\ref{53e}), (\ref{53k}),
and the relations 
$\sum_{n=1}^N p_n^2=\tr \rho^2$, 
$\sum_{n=1}^N p_n^3\leq (\tr \rho^2)^{3/2}$,
derived in Appendix A, as follows
\begin{eqnarray}
\sigma_{\!\! A}^2
& \leq &
K^{(0)} \da^2 \, \tr \rho^2 +
2 K^{(2)} \da^2 \left(2 \, [\tr \rho^2]^{3/2} + [\tr \rho^2]^2\right) \ .
\label{53h}
\end{eqnarray}

Our next goal is to find upper and lower bounds for $G(x)$
for $x>0$ ($x=0$ is trivial) in order to estimate $K^{(k)}$ 
from (\ref{52a}).
To this end, we consider $x>0$ as arbitrary but fixed,
and consider the right hand side in (\ref{30c}) as a function
of ${\bf p}:=(p_1,...,p_N)$,
\begin{eqnarray}
Q({\bf p}) & := & \prod_{n=1}^N (1+xp_n)^{-1} \ .
\label{54a}
\end{eqnarray}
The basic idea is to determine its maximum and the minimum 
under the three constraints (\ref{1b}), (\ref{1c}), and
$p_n\leq p_{max}$ for all $n$, 
where 
\begin{eqnarray}
p_{max} & := & \max_{n}p_n \ .
\label{55a}
\end{eqnarray}
The differential/variation of (\ref{54a}) reads
\begin{eqnarray}
\delta Q({\bf p}) & := & - x\, Q({\bf p}) \sum_{n=1}^N \frac{\delta p_n}{1+xp_n} \ ,
\label{56a}
\end{eqnarray}
complemented by the constraints 
$\sum\delta p_n=0$, 
$\delta p_n\geq 0$ if $p_n=0$, and
$\delta p_n\leq 0$ if $p_n=p_{max}$.
Observing that $x Q({\bf p})>0$ 
on the right hand side of (\ref{56a}) and that the factors 
$1/(1+xp_n)$ are smaller (but still positive) for 
large $p_n$ than for small $p_n$
implies that $Q({\bf p})$ can always be made
larger ($\delta Q({\bf p})>0$) by making the already large
$p_n$ still larger ($\delta p_n>0$) and the already small
$p_n$ still smaller ($\delta p_n<0$).
As a consequence, $Q({\bf p})$ is minimal if all $p_n$ are equal,
implying that
\begin{eqnarray}
G(x) \geq (1+x/N)^{-N} \ .
\label{57a}
\end{eqnarray}
On the other hand, $Q({\bf p})$ cannot be increased any 
more if and only if the small $p_n$ have reached the lower limit 
$p_n=0$ and the large $p_n$ the upper limit $p_n=p_{max}$.
Denoting by $N_{max}$ the number of those $p_n$ equal
to $p_{max}$, their total weight $N_{max}p_{max}$ 
is generically still not exactly equal to unity for 
any integer $N_{max}$. Hence there must remain one weight
$p_n$ with a value $1-N_{max}p_{max}=:p_0\in [0,p_{max}]$ 
in order to fulfill the constraint (\ref{1c}).
All in all, this implies the upper bound
\begin{eqnarray}
G(x) \leq (1+xp_{max})^{-N_{max}} (1+xp_0)^{-1} \ .
\label{58a}
\end{eqnarray}
Next we note that for any $a>0$ the auxiliary
function $f(y):=\ln(1+ay)-y\ln(1+a)$
is zero for $y=0$ and $y=1$ and has a 
negative second derivative for $y\geq 0$, 
implying that $f(y)\geq 0$ for all $y\in[0,1]$.
Setting $a=xp_{max}$ and $y=p_0/p_{max}$
it follows that 
$(1+xp_0)^{p_{max}}\geq (1+xp_{max})^{p_{0}}$
and due to $p_0:=1-N_{max}p_{max}$ that
$1+xp_0\geq (1+xp_{max})^{1/p_{max}-N_{max}}$.
With (\ref{58a}) we thus can infer that
\begin{eqnarray}
G(x) & \leq & (1+xp_{max})^{-1/p_{max}} \ .
\label{59a}
\end{eqnarray}

From (\ref{1c}) and (\ref{55a}) we see that 
\begin{eqnarray}
p_{max} & \geq & 1/N \ .
\label{60a}
\end{eqnarray}
Further, the upper bound (\ref{59a}) yields 
finite integrals (\ref{52a}) only if
\begin{eqnarray}
p_{max}<1/3 \ .
\label{61a}
\end{eqnarray}
Note that this condition implies $N>3$ and hence
condition (\ref{53b}) is automatically satisfied.
Taking for granted (\ref{61a}) inequality 
we can infer by exploiting the bounds (\ref{57a}) 
and (\ref{59a}) in (\ref{52a}) and after performing $k$ partial
integrations that
\begin{eqnarray}
K^{(k)} & = & \prod_{j=1}^{k+1}\frac{1}{1-jp^{(k)}} , \ k=0,1,2
\label{62a}
\\
p^{(k)} & \in & [1/N, p_{max}]
\label{63a}
\end{eqnarray}
With (\ref{53h}) we thus obtain for the variance the upper bound
\begin{eqnarray}
\sigma_{\!\! A}^2
& \leq & \da^2 \, \left(
\frac{\tr \rho^2}{1- p_{max}}+
\frac{4 \, [\tr \rho^2]^{3/2} + 2[\tr \rho^2]^2}
{(1- p_{max})(1- 2p_{max})(1- 3p_{max})}\right) \ .
\label{64a}
\end{eqnarray}

\section{Discussion of the main results}
The upper bound (\ref{64a}) for the variance from
(\ref{20b}) is the first main result of our paper.
In our derivation we have assumed that $\langle A\rangle=0$ 
(see (\ref{20c})), but since the variance from
(\ref{20b}) and also all the other quantities appearing in 
(\ref{64a}) remain unchanged upon replacing $A$ by
$A-\langle A\rangle$ we can conclude that (\ref{64a})
remains valid for arbitrary $A$.
Moreover, we made the assumption that $N<\infty$ and
$p_n>0$ for all $n$ in deriving (\ref{64a}).
Since neither of the quantities appearing in the final 
result (\ref{64a}) give rise to any problem in the
limit $p_n\to 0$, the assumptions $p_n>0$
can be given up as well.
Finally, the limit $N\to\infty$ depends on
the meaning and existence of this limit 
for the quantities appearing on the right 
hand side of (\ref{64a}). 
In particular after dividing 
both sides by $\da^2$ (see below), we expect
that in many important cases this limit will not
give rise to any problems.
The only remaining condition for (\ref{64a})
to be applicable is thus $p_{max}<1/3$ (see (\ref{61a})).

In the special case of a microcanonical density operator
(\ref{1d}) we have obtained as a second main result
the exact relation (\ref{53k1})
for the variance under the same assumptions as above,
namely $\langle A\rangle=0$,  $N<\infty$, and $p_n>0$ 
for all $n$.
Accordingly, for more general observables $A$ with 
$\langle A\rangle\not =0$ we have to replace
$A$ by $A-\langle A\rangle$ in (\ref{53k1}).
Next, if not all $p_n$ are positive and thus equal to
$1/N$, then we have to replace $A$ by $A_+$ and
$N$ by $N_+$, as discussed at the beginning of Sect. III.
All in all, we thus obtain for a microcanonical density 
operator (\ref{1d}) the general exact result
\begin{equation}
\sigma_{\!\! A}^2 
= \frac{N_+}{N_+ +1}\tr[\rho_{mic} (A_+ -\langle A_+\rangle)]^2 =
\frac{1}{N_+ +1}\left(\frac{\tr A_+^2}{N_+}-
\left[\frac{\tr A_+}{N_+}\right]^2\right) \ ,
\label{70a}
\end{equation}
where $A_+$ is the projection of the original
operator $A$ onto the subspace spanned by
the basis vectors $|n\rangle$ with
non-trivial weights $p_n>0$, see Eq. (\ref{8c}).
As before, the meaning and existence of the limit
$N\to\infty$ depends on the behavior of $N_+$, $\tr A_+/N_+$,
and $\tr A_+^2/N_+$ in this limit, but is expected
not to give rise to any problems in many important
cases.

Results similar to (\ref{70a}) have been 
previously derived in Ref. \cite{llo88},
in Ref. \cite{gem04} (see formula (C.17) therein),
and in Ref. \cite{hay07} (see Lemma 3 therein, whose 
proof is very close in spirit to Ref. \cite{gol06a}).
The main difference is that these results only apply to the
special case that $S=\{1,...,N\}$ and 
hence $N_+=N$ in (\ref{1e}), implying
that $\rho_{mic}$ in (\ref{1d}) is proportional to 
the identity operator, cf. (\ref{8b}).
At first glance, a further difference appears to
be that the above mentioned results do not refer to
the GAP measure (\ref{10a}) associated with the above
$\rho_{mic}$ but rather are derived under the assumption
that all (normalized) pure states $|\psi\rangle\in\hr$ are
realized with equal probability.
However, by noting that the latter assumption uniquely
determines the probability density
$p({\bf z})$ for the coefficients $z_n$ in (\ref{4a})
and that the GAP measure does fulfill the assumption
(see below Eq. (\ref{15a2})) we can conclude that 
there is in fact no difference in this respect.
As a by product we can infer that
(\ref{70a}) in particular applies to the case that
all (normalized) pure states $|\psi\rangle$ within the 
subspace $\hr_+\subset\hr$ are realized with equal 
probability and all other $|\psi\rangle\in\hr$ are 
excluded.
After submission of this paper, A. Sugita pointed 
out that the same finding is also contained
in his recent work \cite{sug07}.

Of particular interest in (\ref{64a}) are 
situations for which the bracket on the
right hand side becomes a small quantity.
Therefore, we now focus on the case that the 
so-called purity $\tr\rho^2$ of the
mixed state $\rho$ is low, i.e.
\begin{equation}
\tr  \rho^2  \ll 1 \ .
\label{71a}
\end{equation}
We recall the well-known facts that
the purity is one if and only if $\rho$
corresponds to a pure state
($\rho=|\psi\rangle\langle\psi|$ for some
$|\psi\rangle\in\hr$), is
smaller than unity in any other case,
and takes the minimal possible value 
$1/N$ if $p_n=1/N$ for all $n$ in 
(\ref{1a}).
Roughly speaking, a low purity $\tr \rho^2$  
thus means that the mixed state $\rho$
is very `far' from any 
pure state $|\psi\rangle\in\hr$.

According to Appendix A, the quantity 
$p_{max}$ from (\ref{55a}) can be estimated from
above and from below as follows
\begin{equation}
\tr\rho^2\leq p_{max}\leq\sqrt{\tr\rho^2} \ .
\label{71b}
\end{equation}
Hence, assumption (\ref{71a}) is fulfilled if and only if 
all $p_n$ in (\ref{1a}) are small and is tantamount to the 
condition
\begin{equation}
p_{max}\ll 1 \ .
\label{71d}
\end{equation}
Accordingly, there cannot be just a few
dominating $p_n$ in (\ref{1a}) in the sense
that their sum would already be of 
the order of unity.
In particular, the dimensionality 
$N$ of the Hilbert space $\hr$ must be large
according to (\ref{60a}).

Exploiting (\ref{71a}) and (\ref{71b}) in (\ref{64a}) yields
our final main result
\begin{eqnarray}
\sigma_{\!\! A}^2
& \leq &
\da^2 \, \tr \rho^2 \, (1+\ord(\sqrt{\tr \rho^2}))\ .
\label{80a}
\end{eqnarray}
By rewriting (\ref{70a}) in a form analogous to the
last expression in (\ref{53k1}) one readily sees that
the upper bound (\ref{80a}) is expected to be
quite tight in typical cases.

\section{Summary and Conclusions}
Given any mixed state $\rho$ of low purity (\ref{71a})
there exists at least one probability distribution 
$p({\bf z})$ of pure states (\ref{4a}),
namely the GAP measure (\ref{10a}),
with the following properties:
(i) By randomly sampling pure states $|\psi\rangle$
according to this probability distribution, the
preset statistical ensemble $\rho$ is reproduced.
(ii) Given any observable $A$,
for the overwhelming majority of pure states
$|\psi\rangle$ sampled according to $p({\bf z})$
the expectation value $\langle\psi|A|\psi\rangle$ 
deviates extremely little
from the ensemble averaged expectation value $\tr(\rho A)$ 
compared to the full range $\da$
of {\rm a priori} possible outcomes of
a measurement corresponding to $A$.
The latter statement can be expressed more rigorously 
\cite{rei07} by means of (\ref{80a}) in combination with
Chebyshev's inequality \cite{wik},
and it is tacitly assumed that
this range $\da$ from (\ref{8d}) is
non-zero and remains bounded even in the case 
of an infinite dimensional Hilbert space $\hr$, 
as is the case for any experimentally 
realistic observable $A$ (see Sect. III).

On the one hand, in general
there are other measures $p({\bf z})$
besides the GAP measure which also satisfy
property (i) above but for which
property (ii) may not necessarily
remain true.
On the other hand, $\rho$ fixes all observable
properties of the system via (\ref{2a}), so 
that under typical circumstances any further 
information regarding $p({\bf z})$
is neither necessary nor available.
Hence, in order to uniquely specify $p({\bf z})$ 
for a given $\rho$, one has
either to introduce and justify additional postulates
regarding $p({\bf z})$ \cite{gol06a,pop06}, or to show 
that many or all of the $p({\bf z})$ compatible with $\rho$  
lead to essentially the same final conclusions (ii) \cite{rei07}, 
or one has to include the preparation and 
equilibration process of the system into the 
consideration \cite{simulations}.
In our present work we have focused on the 
first among those three options.

The justification for selecting the 
GAP measure has been discussed
in detail in Ref. \cite{gol06b}.
In particular, it is argued in \cite{gol06b} that
this measure arises naturally when considering
macroscopic systems in thermal equilibrium and
hence is the most appropriate choice, at 
least in cases when $\rho$ is known to
be the canonical density matrix.
Furthermore, as shown in Ref. \cite{gol06b} 
and again in Sect. VI,
this measure is the unique solution in 
the case of a microcanonical density operator
(\ref{1d}) under the additional assumption
that all (normalized) pure states 
$|\psi\rangle\in\hr_+$ are equally likely 
and all other $|\psi\rangle\in\hr$ are excluded,
where $\hr_+$ is the sub Hilbert space 
spanned by all the eigenvectors $|n\rangle$ 
with $p_n>0$ in (\ref{1a}), i.e. the
quantum mechanical analogue of the classical
energy shell within the standard 
microcanonical formalism.

In other words, whenever the Hilbert space $\hr$ 
of the system contains a subspace $\hr_+\subset \hr$ with 
the property that all $|\psi\rangle\in\hr_+$ are 
realized with equal probability and 
all other $|\psi\rangle \in\hr$ are excluded
then the variance $\da$, characterizing the 
dispersion of the random variable
$\langle\psi|A|\psi\rangle$ 
(see (\ref{20b})), is given by the exact relation
(\ref{70a}), where $N_+$ is the dimension of
$\hr_+$ and $A_+$ the restriction/projection
of $A$ to $\hr_+$ (see (\ref{8c})).

\acknowledgments
Special thanks is due Chris Van den Broeck and
Jochen Gemmer for inspiring discussions and to
David Speer for carefully reading the manuscript.

\appendix
\section{}
By means of ({\ref{1a}), the definition of $p_{max}$ in 
(\ref{55a}) and the normalization (\ref{1c}) we
can conclude that
\begin{equation}
\tr  \rho^{k+1} = \sum_{n=1}^N p_n^{k+1} 
\leq\sum_{n=1}^N p_{max}^{k} p_n
= p_{max}^{k}
\label{72a}
\end{equation}
for any integer $k\geq 0$.
For the microcanonical density operator
(\ref{1d}), the above inequality becomes an
equality, i.e. the lower bound 
for $p_{max}$ following from (\ref{72a})
cannot be improved in general.
Likewise, one readily sees that
\begin{equation}
\tr  \rho^{k} = \sum_{n=1}^N p_n^{k} 
\geq \max_{n} p_n^k = p_{max}^{k} \ .
\label{73a}
\end{equation}
Here the inequality becomes an equality
if $p_n\to 1$ for one index $n$ and $p_m\to 0$
for all $m\not =n$, and hence again no
general improvement of the corresponding 
upper bound for $p_{max}$ is possible.
In particular, for $k=2$ we have $p_{max}^2\leq \tr\rho^2$
and hence we can conclude that
\begin{equation}
0\leq p_n^{k} \leq p_{max}^{k}=  (p_{max}^2)^{k/2} \leq (\tr\rho^2)^{k/2}
\label{74a}
\end{equation}
for any integer $k\geq 1$.
Finally, this result yields
\begin{equation}
\tr\rho^k =\sum_{n=1}^N p_n^{k}\leq \sum_{n=1}^N p_{max}^{k-2} p_n^{2} 
= p_{max}^{k-2}\tr\rho^2\leq (\tr\rho^2)^{(k-2)/2}\tr\rho^2 = (\tr\rho^2)^{k/2}
\label{75a}
\end{equation}
for any integer $k\geq 2$.


\end{document}